\documentclass[12pt]{article}
\usepackage{amsmath}
\usepackage{amsfonts,amssymb}

\newcommand{\re}{\mathbb{R}}

\begin{document}

\thispagestyle{empty}

\vspace{.5cm}

\begin{center} {\bf \Large  Nonlinear $W_\infty$ as Asymptotic Symmetry}
\vskip0.3cm
{\bf \Large  of}
\vskip0.3cm
{\bf \Large Three-Dimensional Higher Spin AdS Gravity}

\vspace{1.75cm}

{\large Marc Henneaux} $^{1,2}$ \ \ \& \ \ {\large Soo-Jong Rey} $^{3,4}$

\footnotesize
\vspace{1.25 cm}

${}^1${\em Universit\'e Libre de Bruxelles and International Solvay Institutes}\\
\vspace{0.15cm}
{\em ULB-Campus Plaine CP231, 1050 Brussels, BELGIUM}

\vspace{.3cm}

${}^2${\em Centro de Estudios Cient\'{\i}ficos (CECS), Casilla 1469, Valdivia, CHILE}\\

\vspace{.3cm}

${}^3${\em School of Natural Sciences, Institute for Advanced Study, Princeton NJ 08540 USA}\\

\vspace{.3cm}

${}^4${\em School of Physics and Astronomy \& Center for Theoretical Physics}\\
\vspace{0.15cm}
{\em Seoul National University, Seoul 151-747 KOREA}\\

\end{center}

\vspace {2cm}

\centerline{\bf abstract}
\vspace{1cm}
\noindent
We investigate the asymptotic symmetry algebra of (2+1)-dimensional higher spin, anti-de Sitter gravity. We use the formulation of the theory as a Chern-Simons gauge theory based on the higher spin algebra $hs(1,1)$. Expanding the gauge connection around asymptotically anti-de Sitter spacetime, we specify consistent boundary conditions on the higher spin gauge fields. We then study residual gauge transformation, the corresponding surface terms and their Poisson bracket algebra. We find that the asymptotic symmetry algebra is a nonlinearly realized $W_{\infty}$ algebra with classical central charges. We discuss implications of our results to quantum gravity and to various situations in string theory.

\newpage

\setcounter{page}{1}

\section{Introduction}
\setcounter{equation}{0}
Higher spin (HS) anti-de Sitter (AdS) gravity \cite{Fradkin:1986qy,Fradkin:1987ks,Vasiliev:1992av} is an interesting extension of AdS Einstein-Hilbert gravity, whose various properties turn out to be highly nontrivial compared to the latter.

This HS theory is also expected to be relevant to a variety of situations in string theory. For example, in Maldacena's anti-de Sitter / conformal field theory (AdS/CFT) correspondence
\cite{Maldacena:1997re}, one would like to understand the holographic dual of CFT at weak {}'t Hooft coupling regime. CFTs in this regime are known to possess infinitely many towers of HS currents \cite{wittenTalk}.
By holography, this would mean that the putative closed string dual is at small string tension or large spacetime curvature, and must contain infinitely many towers of HS gauge fields in addition to gravity. One expects that HS AdS gravity is the simplest framework for studying the AdS/CFT correspondence in this regime. In this context, asymptotic symmetry was studied extensively for AdS (black hole) spacetime as the holographic dual of symmetries of CFT at strong {}'t Hooft coupling regime. An interesting question is whether the symmetry persists as the correspondence is interpolated to small {}'t Hooft coupling regime and, if so, how we may identify it as an asymptotic symmetry of the holographic dual, HS AdS gravity.

(2+1)-dimensional AdS gravity is particularly interesting since the theory is simple yet possesses a rich asymptotic symmetry \cite{Brown:1986nw} and provides a concrete framework for studying the AdS/CFT correspondence. It was shown in \cite{Brown:1986nw} that the asymptotic symmetry algebra is the infinite-dimensional conformal algebra in two dimensions, viz. two copies of the Virasoro algebra $Vir \oplus \overline{Vir}$, with central charge
\begin{equation} c = \frac{3\ell}{2G} \ , \label{CC}\end{equation}
where $\ell$ is the anti-de Sitter radius and $G$ is the Newton's constant. Extension to (2+1)-dimensional AdS supergravities \cite{Achucarro:1987vz} was considered in \cite{Banados:1998pi,Henneaux:1999ib}. In this case, the asymptotic symmetry algebra is enlarged to appropriate extended superconformal algebras with quadratic nonlinearities in the currents \cite{Knizhnik:1986wc,Bershadsky:1986ms,Fradkin:1991gj,Fradkin:1992bz,Fradkin:1992km,Bowcock:1992bm}.

The purpose of this work is to report results on the asymptotic symmetry algebra of HS AdS gravity in (2+1) dimensional spacetime. The reason we focus on (2+1) dimensions is because the HS AdS gravity again takes a particularly simple form --- it can be formulated as a Chern-Simons theory based on so-called infinite-dimensional HS algebra $hs(1,1)$ \cite{Blencowe:1988gj,Bergshoeff:1989ns}.  This algebra contains $sl(2,\re) \oplus sl(2,\re)$ as a subalgebra, and hence its Chern-Simons formulation automatically contains three-dimensional AdS gravity \cite{Achucarro:1987vz,Witten:1988hc}.
After briefly reviewing the theory, we provide boundary conditions on the fields that are asymptotically invariant under an infinite-dimensional set of transformations that contains the conformal group at infinity, and whose generators are shown to close according to a classical nonlinear $W_\infty$ algebra.  This algebra is an extension of the classical version of the $W_N$ algebras of \cite{Zamolodchikov:1985wn}.  Classical \cite{Yu:1991bk,Yu:1991cv} and quantum \cite{Bakas:1991fs} nonlinear $W_\infty$ algebras have appeared previously, but unlike the classical $W_\infty$ algebra of \cite{Yu:1991bk,Yu:1991cv}, the asymptotic algebra uncovered here has a nontrivial central charge set by the AdS radius scale measured in unit of the Newton's constant.  In particular, the central charge in the Virasoro subalgebra remains equal to that of the pure gravity (\ref{CC}).

A more detailed presentation of our results as well as supersymmetric extensions will be presented in separate works \cite{ours}.

\section{Higher Spin Anti-de Sitter Gravity}
\setcounter{equation}{0}
We first recapitulate (2+1)-dimensional AdS Einstein-Hilbert gravity coupled to an infinite tower of HS gauge fields. It is well-known that the (2+1)-dimensional AdS Einstein-Hilbert gravity can be reformulated as a Chern-Simons gauge theory with gauge group $SO(2,2)\simeq SO(2,1)\times SO(2,1)$. Following the pioneering work of Blencowe \cite{Blencowe:1988gj}, an approach incorporating the HS gauge fields simply replaces the gauge $SO(2,2)$ gauge group by a suitable infinite-dimensional extension of it. In this work, we shall follow this approach. We should, however, emphasize that our analysis is strictly at classical level and there will make no difference between the Chern-Simons and the Einstein-Hilbert formulations.

The action describing the HS extension is a difference of two Chern-Simons actions \cite{Blencowe:1988gj}:
\begin{equation}
S[\Gamma,\tilde{\Gamma}] = S_{CS}[\Gamma] - S_{CS}[\tilde{\Gamma}] \label{CSAction0}
\end{equation}
where $\Gamma$, $\tilde{\Gamma}$ are connections taking values in the algebra $hs(1,1)$.  This algebra is a 'higher spin algebra' of the class introduced in \cite{Fradkin:1986ka}.  Its properties needed for the present discussion are reviewed in Appendix \ref{AppendixA}, to which we refer for notations and conventions.  The connections contain all HS gauge field components as well as the metric and spin connection. In (\ref{CSAction0}), $S_{CS}$ is the Chern-Simons action, defined by
\begin{equation}
S_{CS}[\Gamma] = \frac{k}{4 \pi} \int_M Tr \left(\Gamma \wedge d \Gamma + \frac{2}{3} \Gamma \wedge \Gamma \wedge \Gamma \right) \ .
\end{equation}
The 3-manifold $M$ is assumed to have topology $\re \times D$ where $D$ is a 2-manifold with at least one boundary on which we shall focus our analysis and  which we refer as 'infinity'. The parameter $k$ is related to the (2+1)-dimensional Newton constant $G$ as $k = \ell/4 G$, where $\ell$ is the AdS radius of curvature.

It is well known that the HS theory (\ref{CSAction0}) embeds the AdS gravity by truncation. Truncating the connections $\Gamma, \tilde{\Gamma}$ to the $sl(2,\re)$ components $A^a, \tilde{A}^a$ and identifying them with the triad and the spin connection:
\begin{equation} A^a_i = \omega^a_i + \frac{1}{\ell} e^a_i \qquad \mbox{and} \qquad \tilde{A}^a_i= \omega^a_i - \frac{1}{\ell} e^a_i,
\end{equation}
one finds that the action takes the form
\begin{eqnarray}
S[\Gamma,\tilde{\Gamma}] = \frac{1}{8 \pi G} \int_M  d^3 x \left(\frac{1}{2} e R + \frac{e}{\ell^2}+ 2{\cal L}_{\rm HS} \right) \ ,
\label{Action1}
\end{eqnarray}
the Einstein-Hilbert gravity with negative cosmological constant. The equations of motion read
\begin{eqnarray}
&& d e^a + \epsilon^{abc} \omega_b \wedge e_c = 0 \nonumber \\
&& d \omega^a + {1 \over 2} \epsilon^{abc} \omega_b \wedge \omega_c + {1 \over 2 \ell^2} \epsilon^{abc} e_b \wedge e_c = 0.
\end{eqnarray}
The last term in (\ref{Action1}) denotes contribution of higher spin fields. For instance, retaining the $sl(3, \re)$ components $A^{ab}, \tilde{A}^{ab}$ as well and identifying them with
\begin{eqnarray}
A_i^{ab} = \omega^{ab}_i + \frac{1}{\ell} e^{ab}_i \qquad \mbox{and}
\qquad
\tilde{A}^{ab}_i = \omega^{ab}_i - \frac{1}{\ell} e^{ab}_i,
\end{eqnarray}
one easily find that the last term in (\ref{Action1}) takes the form
\begin{eqnarray}
{\cal L}_{\rm HS} = \epsilon_{abc} e^a \wedge (\omega^{bd} \wedge \omega^{ce}
+ e^{bd} \wedge e^{ce}) \eta_{de}
+ e^{ab} \wedge (d \omega_{ab} + \epsilon_{dea} \omega^d \wedge {\omega_b}^e) ] \ .
\end{eqnarray}
Equations of motion read
\begin{eqnarray}
&& d e^{ab} + 2 {\epsilon^{ad}}_e (e_d \wedge \omega^{eb} + \omega_d \wedge {e}^{eb}) = 0 \nonumber \\
&& d \omega^{ab} + 2 {\epsilon^{ad}}_e ( \omega_d \wedge \omega^{eb} + {1 \over \ell^2}
e_d \wedge e^{eb}) = 0.
\end{eqnarray}
These are precisely the spin-3 field equations in the background of negative cosmological constant,
expressed in the first-order formalism.
We should, however, note that the HS theory (\ref{CSAction0}) is not a smooth extrapolation of the AdS gravity - for example, integrating out the massless HS gauge fields does not lead to the AdS gravity in any direct and obvious way.

\section{Asymptotic symmetries}
\setcounter{equation}{0}
\subsection{boundary conditions and surface terms}
With Chern-Simons formulation of the (2+1)-dimensional HS AdS gravity at hand, we are ready to study global gauge symmetries at asymptotic infinity. We shall from now on focus on either chiral sector in (\ref{CSAction0}). The analysis for the other chiral sector proceeds in exactly the same way. We shall also work in units of $\ell = 1$, unless otherwise stated.

In the case of the (2+1)-dimensional AdS Einstein-Hilbert gravity, it was shown in \cite{Coussaert:1995zp,Banados:1998pi,Henneaux:1999ib} that the boundary conditions of \cite{Brown:1986nw} describing asymptotically AdS metrics is given in terms of the $sl(2,\re)$ connections of the Chern-Simons formulation by
\begin{equation}
A \sim \Big[- \frac{1}{r} \frac{2 \pi }{k} L(\phi, t) \, X_{11} + r X_{22} \Big] dx^+ - \Big[ \frac{1}{r} \frac{X_{12}}{2}\Big] dr \,
\end{equation}
with the other chirality sector fulfilling a similar condition. Here, $x^\pm$ are chiral coordinates, $x^\pm = t \pm \phi$ and $L$ is an arbitrary function of $t$ and $\phi$. We denoted the $sl(2,\re)$ generators as
$X_{11}, X_{22}, X_{12}$. See appendix A.2 for our conventions and notations.

It is convenient to eliminate the leading $r$-dependence by performing the gauge transformation \cite{Coussaert:1995zp,Banados:1998pi,Henneaux:1999ib}
\begin{eqnarray}
\Gamma_i \rightarrow \Delta_i = \Omega \partial_i \Omega^{-1} + \Omega \Gamma_i \Omega^{-1}, \qquad
\tilde{\Gamma}_i \rightarrow \tilde{\Delta}_i = \Omega \partial_i \Omega^{-1} + \Omega \tilde{\Gamma}_i \Omega^{-1} \ , \label{GaugeTransf}
\end{eqnarray}
where $\Omega$ depends only on $r$ and is given by
\begin{equation}
\Omega = \left(
     \begin{array}{cc}
       r^{\frac{1}{2}} &0 \\
       0 & r^{-\frac{1}{2}} \\
     \end{array}
   \right) \label{FormOfb}.
\end{equation}
In the new connection $\Delta_i$, the only component that does not vanish asymptotically is $\Delta_+ \equiv \Delta$, given by
\begin{equation}
\Delta \sim  X_{22} - \frac{2 \pi}{k} L(\phi, t) \, X_{11}.
\end{equation}
We see that the asymptotic boundary conditions are encoded entirely to the highest-weight component, spanned in the present case by the generator $X_{11}$.

We shall generalize these boundary conditions to HS gauge fields. In the conventions and notations of appendix A.3, we proceed by allowing non-zero components of the $X^{(2s)}$. Intuitively, it suffices to vary only the highest-weight components spanned by the generators whose indices are all $1$, viz. $X^{(2s, 0)}$. Thus, we require that, after gauge transformation (\ref{GaugeTransf}), the connection $\Delta$ behaves asymptotically as
\begin{equation}
\Delta \sim  X_{22} - \frac{2 \pi }{k} L \, X_{11} +12 \frac{2 \pi }{k} M \, X_{1111} + \hbox{``higher"}  \ . \label{AsymptoticForm}
\end{equation}
Here, ``higher" denotes terms involving the generators $X^{(2s, 0)}$ of higher spin $2s$ ($s \geq 3$) and $L$, $M$, ... are arbitrary functions of $t$ and $\phi$.  The numerical factors are chosen to get correct normalization in the gauge functional (\ref{SurfaceTerms}) below.

The boundary conditions are preserved by the residual gauge transformations $\Lambda$
\begin{equation}
\delta \Delta = \Lambda' + [\Delta, \Lambda]  \label{VariationDelta}
\end{equation}
that maintain the behavior at asymptotic infinity.  Here, the prime denotes derivative with respect to $x^+$. Recall that $\Lambda$ does not depend asymptotically on $x^-$ in order to preserve $\Delta_- = 0$ at asymptotic infinity, so the derivative with respect to $x^+$ is also the derivative with respect to $\phi$. As shown in the next subsection, these asymptotic symmetries are spanned by the gauge parameter
\begin{equation}
\Lambda = \varepsilon X_{22} + \sum_{s \geq 2} \eta_{s+1} X^{(0,2s)} + \lambda \ .  \label{AsymptoticGauge}
\end{equation}
Here, $\varepsilon$ and $\eta_{s+1}$ are mutually independent arbitrary functions of $x^+$. Also, $\lambda$ involves only the generators $X^{(p,q)}$ with at least one index equal to $1$ (i.e., $p>0$) and is completely determined through the asymptotic conditions in terms of $\varepsilon$ and $\eta_{s+1}$. The lower order terms in $\lambda$ take the form
\begin{eqnarray}
\lambda &=& \left(\frac{1}{2} \varepsilon'' - \frac{2 \pi}{k} \varepsilon L + a_{(2,0)} \right)X_{11} + \left(\frac{1}{2} \varepsilon' + a_{(1,1)} \right) X_{12}  \nonumber \\
&& + \sum_{p \geq 1, q\geq 0, p+q= 2k \geq 4} A_{(p,q)} X^{(p,q)} \ , \label{AsymptoticGaugeSecond}
\end{eqnarray}
where $a_{(2,0)}$ and $a_{(1,1)}$ are determined by $\eta_{s+1}$'s (independent of $\varepsilon$) and where the coefficients $A_{(p,q)} \ (p+q \ge 4)$ are also completely determined by $\varepsilon$ and $\eta_{s+1}$.

Therefore, we see that the asymptotic symmetries are completely encoded to the independent functions $\varepsilon$ and $\eta_{s+1}$ in the gauge parameter (\ref{AsymptoticGauge}). We stress again that they are arbitrary functions of $x^+$ and thus arbitrary functions of $\phi$ at a given time $t$.

According to the general principle of gauge theory, these asymptotic symmetries are generated in the equal-time Poisson bracket by the spatial integral $G[\Lambda] = \int d^2x \, {\rm Tr} (\Lambda {\mathcal G}) + S_\infty$,  where (i) ${\mathcal G}\equiv ({\mathcal G}_A)$ are the Chern-Simons-Gauss constraints, equal to minus the factor of the temporal components of the connection in the action, and (ii) $S_\infty$ is a boundary term at asymptotic infinity chosen such that the variation $\delta G[\Lambda]$ of the generator $G[\Lambda]$ contains only un-differentiated field variations under the given boundary conditions \cite{Regge:1974zd}. This is the requirement that $G[\Lambda]$ has well-defined functional derivatives.
Applying this procedure and using the fact that the generators $X^{(2s,0)}$ (which are the only ones that appear in $\Delta$ except for $X_{22}$) are paired in the scalar product with $X^{(0,2s)}$, one gets
\begin{equation}
G[\Lambda] = \oint d \phi \ \left( \varepsilon L + \eta M + \cdots \right) \ ,  \label{SurfaceTerms}
\end{equation}
up to bulk terms that vanish on-shell. Here, $\eta$ is abbreviation of $\eta_3$ and the ellipses denote contribution of HS terms involving $\eta_{s+1}$ for $s \ge 3$. The normalization factors in (\ref{AsymptoticForm}) were chosen chosen so as not to have factors in (\ref{SurfaceTerms}). In the next section, we shall show that the asymptotic symmetry generated by $G[\Lambda]$ is a nonlinearly realized $W_\infty$ algebra with classical central charges.

%\subsection{Proof of (\ref{AsymptoticGauge}) and (\ref{AsymptoticGaugeSecond})}
\subsection{general structure of symmetry transformations}
In the previous subsection, we argued that the gauge parameter generating the asymptotic symmetry algebra
takes the form of (\ref{AsymptoticGauge}) where $\lambda$ is given by (\ref{AsymptoticGaugeSecond}). Here, we prove this and further identify the general structure generating the sought-for HS symmetry algebra.

The condition that $\Delta + \delta \Delta$, with $\Delta$ given by
(\ref{AsymptoticForm})
and $\delta \Delta$ given by (\ref{VariationDelta}), should take the same form as $\Delta$ leads to conditions on the gauge parameter $\Lambda$.  This gauge parameter  has a priori the general form
(\ref{AsymptoticGauge}) but with $\lambda$ not yet known.  We want to prove that the conditions on $\Lambda$ yield no restriction on $\varepsilon$ and $\eta_{s+1}$, while completely determine $\lambda$ in terms of $\varepsilon$ and $\eta_{s+1}$.
To that end, we first observe that $\delta \Delta$ reads asymptotically
\begin{equation}
\delta \Delta =   - \frac{2 \pi }{k} \delta L \, X_{11} +12 \frac{2 \pi }{k} \delta M \, X_{1111} + \hbox{``More"} \ ,   \label{DeltaDelta}
\end{equation}
where ``More" denotes terms involving the generators $X^{(2k,0)}$ of higher orders (i.e., $k \geq 3$).  Thus, $\delta \Delta$ involves only the generators $X^{(2k,0)}$ (no index $2$).  We must therefore require that all terms proportional to the generators $X^{(2k)}$ with at least one index equal to $2$ should cancel in  (\ref{VariationDelta}).

To analyze this requirement, it is useful to have a notation that counts the number of indices $1$ and $2$ in the generators. Therefore, we rewrite $\Delta$ as
\begin{eqnarray} \Delta = X_{22} + \sum_{k\geq 1} N_{(2k,0)} X^{(2k,0)}
\end{eqnarray}
where the coefficients $N_{(2,0)}$ and $N_{(4,0)}$ are evidently proportional to $L$ and $M$, respectively. We also rewrite $\Lambda$ as
\begin{eqnarray}  \Lambda &=& \sum_{k \geq 1} \rho_{(0,2k)} X^{(0, 2k)} + \sum_{k \geq 1} \rho_{(1,2k-1)} X^{(1, 2k-1)} + \sum_{k \geq 1} \rho_{(2,2k-2)} X^{(2, 2k-2)} \nonumber \\&+&\sum_{k \geq 2} \rho_{(3,2k-3)} X^{(3, 2k-3)} + \sum_{k \geq 2} \rho_{(4,2k-4)} X^{(4, 2k-4)} \nonumber \\
&+& \sum_{k \geq 3} \rho_{(5,2k-5)} X^{(5, 2k-5)} + \cdots \nonumber \end{eqnarray}
The first term in this expansion is a rewriting of  $\varepsilon X_{22} + \sum_{k \geq 2} \eta_{k+1} X^{(0,2k)}$ part, while $\lambda$ is the sum of all the other terms.

The idea now is to investigate consequences of the requirement that  all terms proportional to the generators $X^{(2k)}$ with at least one index equal to $2$ ought to cancel in  (\ref{VariationDelta}) by examining (i) first the terms containing the generators $X^{(0,2k)}$ with no index equal to 1 in $\Lambda' + [ \Delta, \Lambda]$ (viz. all indices equal to 2), (ii) next those with only one index equal to 1, (iii) next those with only two indices equal to 1, etc.

A simple calculation shows that the coefficient $c_{(0,2k)}$ of $X^{(0,2k)}$
in $\Lambda' + [ \Delta, \Lambda]$ (no indices equal to 1) is given by
\begin{eqnarray}
c_{(0,2k)} \sim \rho_{(0,2k)}' + \rho_{(1,2k-1)} + f_0 (\rho_{(0,2i)},N_{(2j,0)}) \ ,
\end{eqnarray}
As our goal is to explicitly indicate how the structure emerges, we presented the terms only schematically by dropping numerical factors. The term $f_0$ is an infinite sum of bilinears in the $\rho_{(0,2i)}$ 's and the $N_{(2j,0)} $'s.  The first contribution to $f_0$ comes from the bracket of $X_{22}$ with $\sum_{k \geq 1} \rho_{(1,2k-1)} X^{(1, 2k-1)}$ (one $1$ s replaced by one $2$), while the second contribution to $f_0$ arises from the bracket $[\ \sum_{k\geq 1} N_{(2k,0)} X^{(2k,0)} , \sum_{k' \geq 1} \rho_{(0,2k')} X^{(0, 2k')} \ ]$ which is the only bracket in $[\ \sum_{k\geq 1} N_{(2k,0)} X^{(2k,0)} , \Lambda \ ]$ yielding generators $X^{(0,2m)}$ with no index equal to $1$. This bracket yields other generators as well, but they only contribute to the equations at the subsequent levels.  Thus, we can regard the condition $c_{(0,2k)} = 0$ as determining the coefficients $\rho_{(1,2k-1)}$ of $X^{(1, 2k-1)}$ in $\lambda$  in terms of the $\rho_{(0,2i)}$'s and the $N_{(2k,0)}$'s of the connection $\Delta$.

Note that even though $f$ is an infinite sum, there is only a finite number of terms involving a given $\rho_{(0,2k')} $ because one must have $k<k'$ for the bracket $[\ \sum_{k\geq 1} N_{(2k,0)} X^{(2k,0)} , \sum_{k' \geq 1} \rho_{(0,2k')} X^{(0, 2k')} \ ]$ to yield a non vanishing term involving $X^{(0,2m)}$ ($m = k'-k$).  It is also easy to check that  $c_{(0,2)}$ is explicitly given
by
\begin{eqnarray}
c_{(0,2)} = \rho_{(0,2)}' - 2 \rho_{(1,1)} + \hbox{``more"}  \ ,
\end{eqnarray}
where numerical factors are reinstated and ``more" denotes terms independent of $\rho_{(0,2)}$. The condition $c_{(0,2)}= 0$ then implies the expression (\ref{AsymptoticGaugeSecond}) for the coefficient of $X_{12}$ in $\lambda$.

The next step is to examine the coefficient $c_{(1,2k-1)}$ of $X^{(1,2k-1)}$ in $\Lambda' + [ \Delta, \Lambda]$ (only one index equal to $1$). By a similar reasoning, one finds
\begin{eqnarray} c_{(1,2k-1)} \sim \rho_{(1,2k-1)}' + \rho_{(2,2k-2)} + f_1 (\rho_{(0,2i)}, \rho_{(1, 2l-1)}, N_{(2j,0)}).
\end{eqnarray}
Therefore, the requirement $c_{(1,2k-1)} =0$ determines the $\rho_{(2,2k-2)}$'s in terms of  the $\rho_{(0,2i-1)}$'s and the $\rho_{(1,2j-1)}$'s.  Since the $\rho_{(1,2j-1)}$'s are functions of the $\rho_{(0,2i-1)}$'s that have been determined at the previous step, the $\rho_{(2,2k-2)}$'s are determined in terms of  the $\rho_{(0,2i-1)}$'s.

Note again that even though there is an infinite number of terms in $c_{(1,2k-1)}$ because of $f_1$, there is only a finite number of terms containing a given $\rho_{(0,2j)} $.  One finds in particular that $c_{1,1}$ takes the schematic form
\begin{eqnarray}
c_{(1,1)} = \rho_{(1,1)}' -  \rho_{(2,0)} + N_{(2,0)} \rho_{(0,2)}  + \hbox{``more"}
\end{eqnarray}
so that the equation $c_{(0,2)}= 0$ implies the expression (\ref{AsymptoticGaugeSecond}) for the coefficient of $X_{11}$ in $\lambda$.

The triangular pattern of the procedure is now evident and proceeds similarly at the next levels.  One determines in this fashion recursively not only the coefficients $\rho_{(1,2k-1)}$, $\rho_{(2,2k-2)}$ but also $\rho_{(3,2k-3)}$, $\rho_{(4,2k-4)}$, viz. the complete functional form of $\lambda$, in terms of the  coefficients  $\rho_{(0,2j)}$'s, which remain unconstrained.  The procedure terminates once one has imposed the conditions $c_{(2k-1,1)} = 0$. Consequently, there is no condition imposed on $c_{(2k,0)}$.  Rather, the coefficient $c_{(2k,0)}$ determines the variation of the connection $\Delta$ through $\delta N_{(2k,0)} = c_{(2k,0)}$.  Notice that the procedure introduces nonlinearities through the $f_k$'s.

We have thus established that the gauge parameter generating asymptotic symmetry takes precisely the form given in (\ref{AsymptoticGauge}) and (\ref{AsymptoticGaugeSecond}).

%%%%%%%%%%%%%%%%%%%%%%%%%%%%%%%%%%%%%%%%%%%%%%%%%%%%%%%%%%%%%%%%%%%%%%%%%%%%%%%%%%%%%%%%%%%%%%%%%%%%%%%
\section{Nonlinear $W_\infty$ Symmetry Algebra}
\setcounter{equation}{0}

As we explained above, the variations of the coefficients $N_{(2k,0)}$ of the connection $\Delta$ under the asymptotic symmetries are given by the equation
\begin{eqnarray}
\delta N_{(2k,0)} = c_{(2k,0)} \ , \label{Nvariation}
\end{eqnarray}
where the $c_{(2k,0)}$ are the unconstrained coefficients of the generator $X^{(2k,0)}$ in $\Lambda' + [ \Delta, \Lambda]$.  The recursive method explained in the previous section enables to determine these coefficients in terms of the independent parameters $\rho_{(0,2j)}$'s parametrizing the asymptotic symmetry.

We have recalled in the previous section that the coefficients $N_{(2k,0)}$ of the connection $\Delta$ are themselves the generators of the gauge transformations and hence of the asymptotic symmetries.  In fact, in more compact notations, (\ref{SurfaceTerms}) has the form
\begin{equation}
G[\Lambda] = \oint d \phi \Big( \sum_{k\geq 1}\rho_{(0,2k)} N_{(2k,0)} \Big) \label{SurfaceTerms2}
\end{equation}
up to bulk terms that vanish on-shell. Again, for clarity, we kept the expression schematic regarding normalization of the generators. They will not affect foregoing argument and result. (We shall work out the normalization explicitly in the next section for the truncation to $k \le 2$).

In general, the variation $\delta {\cal O}$ of any phase-space function ${\cal O}$ under the gauge transformation with parameter $\Lambda$  is equal to $\{{\cal O}, G[\Lambda]\}_{\rm PB}$ where $\{, \}_{\rm PB}$ is classical Poisson bracket. Thus, in the present case, we have
\begin{equation}
\delta N_{(2k,0)} = \Big\{ N_{(2k,0)}(\phi),  \int d \phi' \big(\sum_{m \geq 1} \rho_{(0,2m)}(\phi') N_{(2m,0)}(\phi') \big) \Big\}_{\rm PB} \ .
\end{equation}
This observation enables us to read the Poisson bracket commutators of the $N_{(2k,0)}$'s in (\ref{SurfaceTerms2}) among themselves from their  variations (\ref{Nvariation}) \footnote{If one drops the bulk terms as can be done by fixing the gauge in the bulk, the Poisson bracket in question is the corresponding Dirac bracket.  The form of the symmetry algebra does not depend on how one fixes the gauge because the generators are first class.}:
\begin{eqnarray}
\Big\{ N_{(2k,0)}(\phi),  \int d \phi' \big(\sum_{m \geq 1} \rho_{(0,2m)}(\phi') N_{(2m,0)}(\phi') \big) \Big\}_{\rm PB}   = c_{(2k,0)} (\phi)  \ ,
\end{eqnarray}
where we have made it explicit for the angular dependence at a fixed time.  By identifying the coefficients of the arbitrary parameter $\rho_{(0,2m)}$ on both sides of this equation, one can read off the Poisson brackets
\begin{eqnarray}
\big\{ N_{(2k,0)}(\phi), N_{(2m,0)}(\phi') \big\}_{\rm PB}
\end{eqnarray}
and resulting algebra ${\mathcal W}$. In the rest of this section, we sketch the general procedure of extracting ${\cal W}$. To illustrate the procedure concretely, in the next section, we will work out the case corresponding to the truncation of $k,m,... \le 2$.

It is evident from the above analysis that the expression obtained for $\{ N_{(2k,0)}(\phi), N_{(2m,0)}(\phi') \}_{\rm PB}$ is closed, in the sense that it is expressed entirely in terms of the $N_{(2j,0)}$'s.  Terms that are generated from the Poisson bracket $\{ N_{(2k,0)}(\phi), N_{(2m,0)}(\phi') \}_{\rm PB}= \cdots$ are in fact {\sl nonlinear} polynomials in the $N_{(2j,0)}$'s. Therefore, the resulting gauge algebra is not a Lie-type but a nonlinear realization thereof. Furthermore, by construction, the Jacobi identity holds for $\{ N_{(2k,0)}(\phi), N_{(2m,0)}(\phi') \}_{\rm PB}$ because it always holds for the Poisson brackets or the corresponding Dirac brackets after the bulk terms are gauge-fixed.

We claim that the resulting algebra ${\mathcal W}$ is a classical, nonlinearly realized $W_\infty$ with classical central charges.  It is a classical algebra because we are using the Poisson-Dirac bracket of classical quantities and not the commutator of corresponding operators. It also has nontrivial classical central charges.

To support this claim, it suffices to prove that
\begin{enumerate}
\item The algebra ${\mathcal W}$ contains the Virasoro algebra at lowest degree $j=1$, viz. the generators $L \sim N_{(2,0)}$ form a Virasoro algebra with central charge $k/4 \pi$:
\begin{equation}
\{L(\phi), L(\phi')\}_{\rm PB} = -\frac{k}{4 \pi} \partial_\phi^3 \delta(\phi - \phi') + \left(L(\phi) + L(\phi') \right) \partial_\phi \delta(\phi - \phi') \label{LLGen}
\end{equation}
\item The generators $M_{j+1} \sim N_{(2j,0)}$ have the conformal weight $(j+1)$:
\begin{equation}
\{L(\phi), M_{j+1}(\phi')\}_{\rm PB}  =  \left( M_{j+1}(\phi) + j M_{j+1}(\phi') \right) \partial_\phi \delta(\phi - \phi') \ . \label{LMGen}
\end{equation}
\end{enumerate}
To establish these statements, we pick up the terms proportional to $\varepsilon$ in $\delta L$ and $\delta M_{j+1}$.  This is done by first determining the form of $\Lambda$ in the particular case when the only non-vanishing free parameter $\rho_{(0,2j)}$ is $\rho_{(0,2)} \equiv \varepsilon$.  In that case, the solution $\Lambda$ is easily determined to be
\begin{equation}
\Lambda = \varepsilon X_{22} + \frac{1}{2} \varepsilon'  X_{12} + \left(\frac{1}{2} \varepsilon'' - \frac{2 \pi}{k} \varepsilon L  \right)X_{11} + \varepsilon  \sum_{j\geq 2} N_{(2j,0)} X^{(2j,0)} \ , \label{Lambda}
\end{equation}
since with this $\Lambda$, the expression $\Lambda' + [\Delta, \Lambda]$ contains only generators $X^{(2m,0)}$.

The coefficients of the generators $X^{(2m,0)}$ in $\Lambda' + [\Delta, \Lambda]$ give furthermore the variations of $L$ and $N_{(2j,0)}$ ($j>1$).  These are easily derived from (\ref{Lambda}) using
\begin{eqnarray} [\ X^{(2j,0)},X_{12} \ ] = (2j) X^{(2j,0)} \ .
\end{eqnarray}
Explicitly, they read
\begin{eqnarray}
\delta L &=&  -\frac{k}{4 \pi} \varepsilon''' + (\varepsilon L)' +  \varepsilon' L \\
\delta N_{(2j,0)} &=&   (\varepsilon N_{(2j,0)})' + j \varepsilon' N_{(2j,0)} \; \; \; (j>1).
\end{eqnarray}
The relations (\ref{LLGen}) and (\ref{LMGen}) follow immediately from these.

Explicit form of the Poisson-Dirac brackets of the resulting $W_\infty$ algebra and classical central charges therein are obtainable by straightforward though tedious computations.

%%%%%%%%%%%%%%%%%%%%%%%%%%%%%%%%%%%%%%%%%%%%%%%%%%%%%%%%%%%%%%%%%%%%%%%%%%%%%%%%%%%%%%%%%%%%%%%%%%%%%%%%
\section{Truncation to $W_3$ Algebra}
\setcounter{equation}{0}

To illustrate the above procedure explicitly, we truncate the theory by assuming that  all the generators $N_{(2j,0)}$ with $j>2$ are zero. i.e., we keep only $L$ and $M$.   This amounts to truncating the HS algebra by keeping only $X_{\alpha \beta}$, $X_{\alpha \beta \gamma \delta}$ and setting all the other generators to zero. As shown in appendix \ref{AppendixB}, this is a unique, consistent truncation as the Jacobi identity remains to hold. The resulting algebra is $sl(3,\re)$, albeit not in a Chevalley-Serre basis~\footnote{The relation between the Chern-Simons formulation and the symmetric tensor formulation is implicit in \cite{Blencowe:1988gj} (using the vielbein/spin connection-like Vasiliev formulation of higher spins), and underlies the fact that \cite{Blencowe:1988gj} is a theory of higher spins coupled to gravity. Truncating this general relation to $sl(3)$, one gets the 'metric + 3-index symmetric tensor' formulation of the coupled `spin-2 + spin-3' system. This was briefly recalled at the end of section 2.}. We now show that $L$ and $M$ fulfill the classical nonlinear $W_3$ algebra with classical central charges.

The condition that $\Delta + \delta \Delta$, with $\Delta$ given by
\begin{equation}
\Delta \sim  X_{22} - \frac{2 \pi }{k} L \, X_{11} +12 \frac{2 \pi }{k} M \, X_{1111}  \label{AsymptoticFormW3}
\end{equation}
and $\delta \Delta$ given by (\ref{VariationDelta}), should take the same form as $\Delta$ leads to conditions on the coefficients of the gauge parameter $\Lambda$ in the expansion
\begin{equation}
\Lambda = a X_{11} + bX_{12} + \varepsilon X_{22} + m X_{1111} + n X_{1112} + p X_{1122} + q X_{1222} + \eta X_{2222}
\end{equation}
These conditions are explicitly  that $a, b$ are determined as
\begin{equation}
b= \frac{1}{2} \varepsilon ', \qquad a = \frac{1}{2} \varepsilon '' - \frac{2 \pi}{k}  \varepsilon L - 2 \frac{2 \pi}{k} \eta M \ .
\end{equation}
and that $m, n, p, q$ are determined as
\begin{eqnarray}
m &=& \frac{1}{24} \eta'''' - \frac{1}{6}\cdot \frac{2 \pi}{k} (\eta L)'' - \frac{1}{4} \cdot \frac{2 \pi}{k} (\eta' L)' \nonumber \\
&& - \frac{2 \pi}{k} \left(\frac{1}{4} \eta'' - \frac{2 \pi}{k} \eta L\right) L + 12 \frac{2 \pi}{k} \varepsilon M \nonumber \\
n &=& \frac{1}{24} \eta''' - \frac{1}{6} \cdot\frac{2 \pi}{k} (\eta L)' - \frac{1}{4}\cdot \frac{2 \pi}{k} \eta' L \nonumber \nonumber \\
p &=& \frac{1}{12} \eta'' - \frac{1}{3} \cdot \frac{2 \pi}{k} \eta L \nonumber \\
q &=& \frac{1}{4} \eta' \ .
\end{eqnarray}
One also obtains the gauge variations of $L$ and $M$ as
\begin{equation}
\delta L = - \frac{k}{4 \pi} \varepsilon ''' + (L \varepsilon)' + \varepsilon' L + 2 (\eta M)' + \eta' M
\end{equation}
and
\begin{eqnarray}
\delta M &=& \frac{1}{288} \cdot\frac{k}{2 \pi} \eta''''' - \frac{1}{72} (\eta L)''' - \frac{1}{48} (\eta' L)'' \nonumber \\
&& - \frac{1}{12} \Big(\big(\frac{1}{4} \eta'' -\frac{2 \pi}{k}  \eta L\big) L \Big)' \nonumber \\
&& - \frac{1}{12} \Big(\frac{1}{6} \eta'''- \frac{2}{3}\cdot \frac{2 \pi}{k} (\eta L)' - \frac{2 \pi}{k} \eta' L \Big) L\nonumber  \\
& & +  (\varepsilon M)' + 2 \varepsilon' M \ .
\end{eqnarray}

Now, as already recalled above in the general case,  the variation $\delta {\cal O}$ of any phase space function ${\cal O}$ under the gauge transformation with parameter $\Lambda$  is equal to $\{{\cal O}, G[\Lambda]\}_{\rm PB}$ where $\{, \}_{\rm PB}$ is the classical Poisson bracket. One can use this to find the Poisson brackets of $L$ and $M$ from their variations, taking (\ref{SurfaceTerms}) into account.

One finds explicitly that
\begin{eqnarray}
&&\{L(\phi), L(\phi')\}_{\rm PB} = -\frac{k}{4 \pi} \partial_\phi^3 \delta(\phi - \phi') + \big(L(\phi) + L(\phi') \big) \partial_\phi \delta(\phi - \phi') \nonumber \\
&&\{L(\phi), M(\phi')\}_{\rm PB}  =  \big( M(\phi) + 2 M(\phi') \big) \partial_\phi\delta(\phi - \phi')
\nonumber \\
&&\{M(\phi), M(\phi')\}_{\rm PB} = \frac{1}{288}\cdot \frac{k}{2 \pi} \partial_\phi^5 \delta(\phi - \phi') - \frac{5}{144} \big(L(\phi) + L(\phi')\big)\partial_\phi^3 \delta(\phi - \phi') \hspace{1.5cm} \nonumber \\
&& \hspace{3.6cm} + \frac{1}{48} \big( L''(\phi) + L''(\phi') \big) \partial_\phi \delta(\phi - \phi') \nonumber \\
&& \hspace{3.6cm} + \frac{1}{9}\cdot \frac{2 \pi}{k} \big(L^2(\phi) + L^2(\phi') \big) \partial_\phi \delta(\phi - \phi') \ . \nonumber
\end{eqnarray}
This is the classical $W_3$ algebra studied previously in various different contexts \cite{Zamolodchikov:1985wn},
\cite{Mathieu:1988pm}, \cite{Bakas:1989mx}.
%[If one redefines $M \rightarrow \sqrt{12} M$ and set $\frac{k}{2 \pi} = - 4$, one exactly gets the $W_3$ algebra as written in

Upon Fourier mode decomposition, the nonlinear $W_3$ algebra is given by \cite{Mathieu:1988pm}
\begin{eqnarray}
i \big[ \  L_m, L_n \ \big] &=& (n-m) L_{m+n} + {c \over 12} m (m^2 - 1) \delta_{m+n, 0}  \\
i \big[ \ L_m,  V_n \ \big] &=& (2m - n) V_{m+n} \nonumber \\
i \big[ \ V_m,  V_n \ \big] &=& {c \over 360} m (m^2 -1) (m^2 - 4) \delta_{m+n,0} + {16 \over 5 c} (m-n) \Lambda_{m+n}
\nonumber \\
&+& (m-n) \Big( {1 \over 15} (m+n+2) (m+n+3) -{1 \over 6} (m+2) (n+2) \Big) L_{m+n} \ \ \
\nonumber
\end{eqnarray}
where
\begin{eqnarray}
\Lambda_m = \sum_{n=-\infty}^{+\infty}  L_{m-n} \ L_n \ .
\end{eqnarray}
sums quadratic nonlinear terms.
%\begin{eqnarray}
%d_{2n} = -{1 \over 2} (n^2 - 1), \qquad d_{2n+1} = -{1 \over 5} (n-1) (n+2).
%\end{eqnarray}
%
The classical central charges are given by $c = {3 \ell / 2 G}$. The quantum counterpart of this $W_3$ algebra was studied by Zamolodchikov \cite{Zamolodchikov:1985wn} in a different context. Comparing it
with the above classical algebra, one sees that the quantum effects enter to regularization of the quadratic nonlinear terms $\Lambda_m$'s and to the shift $5c \rightarrow 5c + 22$ of the overall coefficient of the quadratic terms. This fits with the fact that classical limit takes $c \rightarrow \infty$.

\section{Discussions}

In this paper, we have established that the asymptotic symmetries of the HS AdS gravity form a nonlinear $W_\infty$ algebra. A salient feature of the emerging classical algebra is that it is determined in a unique manner from the gauge algebra $hs(1,1)$ and the Chern-Simons parameter $k = \ell / 4 G$, without an extra free parameter.  Moreover, this classical algebra has from the outset definite nontrivial central charges expressed solely in terms of the AdS radius and the Newton's constant (and nothing else). In particular, the central charge appearing in the Virasoro subalgebra is just the AdS central charge (\ref{CC}).

Truncation of the higher spin gauge algebra $hs(1,1)$ up to a finite spin $s \le N$ is inconsistent when $N>2$, since the corresponding generators do not form a subalgebra (except when $N=2$). The Poisson-Dirac commutators of $X^{(2j)}$ with $X^{(2j')}$ involve indeed generators $X^{(2(j+j'-1))}$ of degree $2(j+j'-1)$, which is strictly higher than $2j$ and $2j'$ when $j>1$, $j'>1$.  One may try to ignore these higher degree terms but this brutal truncation yields commutators that do not fulfill the Jacobi identity (except for $N=3$ as we pointed out, see appendix \ref{AppendixA}) and so this cannot be done (except for $N= 3$)\footnote{That the case $N=3$ works is somewhat unanticipated and should be considered exceptional from the $hs(1,1)$ point of view.}. On the other hand, from the purely algebraic viewpoint, one might opt to start from the $W_N$ algebra with finite $N$ obtained from algebra of $sl(N)$ gauge invariance and take the limit $N \rightarrow \infty$ to obtain a universal $W$-algebra. However, these two approaches are completely different in spirit since in general the truncation of $hs(1,1)$ up to a finite spin $N$ does not yield the algebra $sl(N)$ for any finite $N>2$.

Nonlinearity of the Poisson-Dirac brackets or commutation relations (compared to the Lie-type $Vir = W_2$  algebra) is an important and distinguishing characteristic of the $W_N$ operator algebra for spin $N>2$. However, in the usual large-$N$ limit, this nonlinearity is typically lost \cite{Bakas:1989xu}, \cite{Bakas:1990sh}, \cite{Pope:1989ew}, \cite{Pope:1990be}:  the resulting $W_\infty$ algebra is usually linear (and also in some cases the classical central charge is absent). Our approach obtains the $W_\infty$ gauge algebra in a completely different way, and in particular does not rely on such a limiting procedure. We note that the nonlinearity of the algebra puts strong constraints through the Jacobi identities. Thus, the nonlinear $W_\infty$ algebra derived in this work,  which is inherent to the $hs(1,1)$-based HS extension of the (2+1)-dimensional AdS Einstein-Hilbert gravity -- in the sense that it  uniquely determined by it -- has a rich and interesting structure. More detailed analysis of this algebraic structure, extensions to supersymmetry and inclusion of spin-1 currents will be reported elsewhere \cite{ours}.

Related to this, it has been known previously that the linear version of the $W_\infty$ algebra is related to the first Hamiltonian structure of the KP hierarchy \cite{Yu:1991ng}. In our nonlinearly realized version of the $W_\infty$ algebra, we speculate the relation goes to the second Hamiltonian structure of the KP hierarchy, the structure proposed by Dickey \cite{Dickey2ndH} from generalizing the Gelfand-Dickey brackets \cite{Gelfand:1975rn} to pseudo-differential operators. It is an interesting question what these relations tell us about the spectrum of classical solutions of the HS AdS gravity.

It is tempting to interpret that the presence of the  $W_\infty$ algebra at infinity implies that the classical solutions of the (2+1)-dimensional HS AdS gravity are labeled by infinitely many conserved charges, among which mass and angular momentum are just the first two. If the interpretation is correct, we expect that these charges play a central role in understanding microstates responsible for the black hole entropy in the regime where the spacetime curvature is large or, in string theory context, the string scale is very low\footnote{One should also mention here the intriguing appearance of the linear $W_\infty$ algebra found in \cite{Bonora:2008he,Bonora:2008nk} in the context of black holes and Hawking radiation.}. A possible holographic dual in this regime was explored recently by Witten \cite{Witten:2007kt} for pure AdS gravity. There, an indication was found that two-dimensional CFT duals are the monster theory of Frenkel, Lepowsky and Meurman or discrete series extensions thereof. Once embedded to string theory, one expects this regime must includes (nearly) massless HS gauge fields in addition to the gravity. This brings in a host of intriguing questions: Are there HS extensions of the monster theory and, if so, what are they? Can the extension be related or interpreted physically to condensation of long strings?

In the context of string theory, the massless HS gauge fields were interpreted to arise via a sort of inverse Higgs mechanism in the limit of vanishing string mass scale (viz. string tension) \cite{Sagnotti:2003qa}. If so, the HS gauge fields would become massive at large but finite string mass scale \cite{Sagnotti:2010at}. In CFT dual, this would be reflected to anomalous violation of conservation laws of the HS currents. Nevertheless, the $W_\infty$ symmetry algebra discovered in this work would be an approximate symmetry  of the CFT duals and should still be useful for understanding these theories.

In addition to the weak {}' t Hooft coupling regime alluded in the Introduction, there is another situation
in string theory where the result of this paper may be applicable. The near-horizon geometry of the small black strings carrying one or two charges is singular in Einstein-Hilbert gravity. One expects that, by the stretched horizon mechanism \cite{Sen:2004dp}, string corrections resolve it to the (2+1)-dimensional AdS spacetime times a compact 7-dimensional manifold $M_7$ characterizing the black string horizon with residual chiral supersymmetries. \cite{Park:1998un}. A concrete suggestion like this was put forward for the 'stretched horizon', near-horizon geometry of the macroscopic Type II and heterotic strings \cite{Dabholkar:2007gp}, \cite{Lapan:2007jx}, \cite{Kraus:2007vu}. In both cases, the near-horizon geometry has curvature radius of order the string scale. So, not just the massless but also all HS string states are equally important for finite energy excitations. This suggests that (2+1)-dimensional HS AdS supergravity theories are appropriate frameworks. It would then be very interesting to identify the origin of the $W_\infty$ symmetry algebra as well as the classical central charges associated with HS currents from the macroscopic superstring viewpoint.

On a more speculative side, our result may also find a potentially novel connection of the (2+1)-dimensional HS AdS gravity to higher-dimensional gravity. It has been known \cite{Takasaki:1984tp}, \cite{Park:1989vq}, \cite{Hu:1994dk} that 4-dimensional self-dual gravity is equivalent to a large $N$ limit of 2-dimensional nonlinear sigma model with Wess-Zumino terms only. The self-dual sector has an infinite-dimensional symmetry algebra which includes the $W_\infty$ algebra. This hints that (2+1)-dimensional HS AdS gravity might be 'holographically dual' to 4-dimensional self-dual gravity, providing a concrete example of heretofore unexplored gravity-gravity correspondence.

Centered to all these issues, the most outstanding question posed by our work is: \hfill\break
\centerline{\sl What are the black holes carrying $W_\infty$ hairs in HS AdS gravity?}
\vskip0.45cm
We are currently exploring these issues and intend to report progress elsewhere.

\section*{Acknowledgement}
We thank Nima Arkani-Hamed, Juan Maldacena and Edward Witten for useful discussions.
MH is grateful to the Institute for Advanced Study (Princeton) for hospitality during this work and to the Max-Planck-Institut f\"ur Gravitationphysik (Potsdam) where it was completed. SJR is grateful to the Max-Planck-Institut f\"ur Gravitationphysik (Potsdam) during this work and to the Institute for Advanced Study (Princeton) where it was completed. We both acknowledge support from the Alexander von Humboldt Foundation through a Humboldt Research Award (MH) and a Bessel Research Award (SJR). The work of MH is partially supported by IISN - Belgium (conventions 4.4511.06 and 4.4514.08), by the Belgian Federal Science Policy Office through the Interuniversity Attraction Pole P6/11 and by an ARC research grant 2010-2015. The work of SJR is partially supported in part by the National Research Foundation of Korea grants KRF-2005-084-C00003, KRF-2010-220-C00003, KOSEF-2009-008-0372, EU-FP Marie Curie Training Program (KICOS-2009-06318), and the U.S. Department of Energy grant DE-FG02-90ER40542.

\appendix

\section{Higher Spin Algebra}
\setcounter{equation}{0}
\label{AppendixA}

\subsection{Definition}
The higher spin algebra $\mathcal{A}$ in (2+1)-dimensional spacetime is the direct sum of two chiral copies of $hs(1,1)$:
\begin{equation}
\mathcal{A} = hs(1,1)_L \oplus hs(1,1)_R
\end{equation}
The infinite-dimensional algebra $hs(1,1)$ itself is defined as follows. Consider an auxiliary space ${\mathcal P}$ of polynomials of even degree in two commuting spinors $\xi^1$, $\xi^2$. One defines the trace of a polynomial $f \in {\mathcal P}$ as
\begin{equation}
\mbox{Tr} \, f = 2f(0) \equiv 2f(\xi)\Big\vert_{\xi=0}.
\end{equation}
Here, the factor 2 is included to match the traces of $(2 \times 2)$ matrices considered below. Then, the elements of the  algebra $hs(1,1)$  are the elements of ${\mathcal P}$ with no constant term, viz. traceless polynomials.

To define the Lie bracket, one first considers the star-product defined by
\begin{equation}
(f \star g)(\xi) = \exp\Big[ i\Big(\frac{\partial}{\partial \eta^1}\frac{\partial}{\partial \zeta^2} - \frac{\partial}{\partial \eta^2}\frac{\partial}{\partial \zeta^1}\Big)\Big] \ f(\eta)g(\zeta)\Big\vert_{\eta=\zeta=\xi}
\label{DefStar}\end{equation}
The star-product is associative. Although non-commutative, the star-product is trace-commutative:
\begin{equation}
(f \star g)(0) = (g \star f)(0), \label{Symm0}
\end{equation}
viz.
\begin{equation}
\mbox{Tr}(f\star g) = \mbox{Tr}(g\star f) \label{Symm}
\end{equation}
because $f$ and $g$ are polynomials of even degree.  The Lie bracket in the algebra $hs(1,1)$ is just the $\star$-commutator (modulo a numerical factor chosen for convenience to be $\frac{1}{2i}$):
\begin{equation}
[f,g] \equiv \frac{1}{2i}\left(f \star g - g \star f\right) = \sin \Big(\frac{\partial}{\partial \eta^1}\frac{\partial}{\partial \zeta^2} - \frac{\partial}{\partial \eta^2}\frac{\partial}{\partial \zeta^1}\Big) \, f(\eta)g(\zeta)\Big\vert_{\eta=\zeta=\xi}.
\end{equation}
It fulfills the Jacobi identity because the star-product is associative.

The Lie algebra $hs(1,1)$ possesses a symmetric and invariant bilinear form denoted $(,)$, defined by
\begin{equation}
(f,g) \equiv \mbox{Tr}(f\star g).
\end{equation}
This bilinear form is symmetric because of the trace-commutativity (\ref{Symm}) and invariant
\begin{equation}
(f, [g,h]) = ([f,g], h)
\end{equation}
because of the associativity of the star-product and (\ref{Symm}) again.  The invariant symmetric bilinear form is non-degenerate.

\subsection{$sl(2,\re)$ subalgebra}
The polynomials of degree 2 form a subalgebra isomorphic to $sl(2,\re)$.  Taking as a basis of this subspace as
\begin{equation}
X_{11} = \frac{1}{2}( \xi^1)^2, \qquad X_{12} = \xi^1 \xi^2, \qquad X_{22} =  \frac{1}{2}( \xi^2)^2 \ ,
\end{equation}
one finds
\begin{equation}
[X_{11}, X_{12}] = 2 X_{11}, \; \; [X_{11}, X_{22}] = X_{12}, \; \; [X_{12}, X_{22}] = 2 X_{22} \ .
\end{equation}
One can thus identify the $X_{(\alpha \beta)}$ with the standard Chevalley-Serre generators $\{h,e,f\}$ as follows: $X_{12} = - h$, $X_{11}= -e$ and $X_{22}= f$.

Moreover, traces of the products of $X_{(\alpha \beta)}$'s match with traces of the products of the corresponding $(2 \times 2)$ matrices. The non-zero scalar products are
\begin{equation}
(X_{12},X_{12}) = 2, \qquad (X_{11}, X_{22}) = - 1, \qquad (X_{22}, X_{11}) = -1.
\end{equation}

The subalgebra $hs(1,1)$ splits into a direct sum of representations of $sl(2,\re)$:
\begin{equation}
hs(1,1) = {\Large\oplus}_{k \geq 1} D_k \ ,
\end{equation}
where the spin $k$ representation $D_k$ corresponds to the homogeneous polynomials of degree $2k$.  The trivial representation $D_0$ does not appear because we consider traceless polynomials.
 %(no constant piece).
It is straightforward to verify that the subspaces $D_k$ and $D_{k'}$ are orthogonal for $k \not= k'$ and that the scalar product is non-degenerate on each $D_k$.

We emphasize that, as showed in the text, {\em the representation $D_k$ yields asymptotically the generators $M_{k+1}$ of conformal spin $k+1$.} Notice the shift of the spin label by one unit.

\subsection{more commutation relations}
We list here the commutation relations involving $D_1$ and $D_2$.
A basis of the representation $D_2$ of $hs(1,1)$, polynomials of order 4, may be taken to be
\begin{eqnarray}
X_{1111} = \frac{1}{4!} (\xi^1)^4, \qquad  X_{1112} = \frac{1}{3!} (\xi^1)^3 \xi^2, \qquad X_{1122} = \frac{1}{4} (\xi^1)^2 (\xi^2)^2, \nonumber
\end{eqnarray}
\begin{eqnarray}
X_{1222} = \frac{1}{3!} \xi^1 (\xi^2)^3, \qquad X_{2222} = \frac{1}{4!}  (\xi^2)^4 \ .
\nonumber
\end{eqnarray}
More generally, we define
\begin{equation}
X^{(p,q)} \equiv X_{\underbrace{1 \cdots 1}_p \underbrace{2 \cdots 2}_q}  = \frac{1}{p!} \frac{1}{q!} \left( \xi^1\right)^p \left(\xi^2 \right)^q  \; \; \; \; \hbox{ with $p + q$ even}
\end{equation}
The vectors $X^{(p,q)}$ with $p+q = 2k$ form a basis of $D_k$.  We use the collective notation $X^{(2k)}$ for the $X^{(p,q)}$'s with $p+q = 2k$.

The brackets of the $X^{(4)}$'s with the $X^{(2)}$'s are given by
\begin{eqnarray}
&&Ê[X_{11}, X_{1111}] = 0, \qquad [X_{11}, X_{1112}] = +4 X_{1111}, \qquad [X_{11}, X_{1122}] = +3 X_{1112},\nonumber \\
&& [X_{11}, X_{1222}] = +2 X_{1122}, \qquad [X_{11}, X_{2222}] =  +X_{1222},\nonumber \\
&&Ê[X_{12}, X_{1111}] = -4 X_{1111}, \qquad [X_{12}, X_{1112}] = -2 X_{1112}, \qquad [X_{12}, X_{1122}] = 0, \nonumber \\
&& [X_{12}, X_{1222}] = +2 X_{1222}, \qquad [X_{12}, X_{2222}] =  +4 X_{2222},\nonumber \\
&&Ê[X_{22}, X_{1111}] = - X_{1112}, \qquad [X_{22}, X_{1112}] = -2 X_{1122}, \nonumber \\
&& [X_{22}, X_{1122}] = -3 X_{1222}, \qquad
[X_{22}, X_{1222}] = -4 X_{2222}, \qquad [X_{22}, X_{2222}] =  0, \nonumber
\end{eqnarray}

The brackets between $X^{(4)}$'s are
\begin{eqnarray}
&&[X_{1111}, X_{1112}] = 20 \, X_{111111}, \qquad [X_{1111}, X_{1122}] = 10 \, X_{111112}, \nonumber \\
&& [X_{1111}, X_{1222}] = 4 \, X_{111122} - \frac{1}{3} X_{11}, \nonumber \\
&& [X_{1111}, X_{2222}] =  X_{111222} - \frac{1}{3!} X_{12}, \nonumber \\
&&[X_{1112}, X_{1122}]= 8 \,  X_{111122} + X_{11}, \nonumber \\
&& [X_{1112}, X_{1222}] = 8 \,  X_{111222} + \frac{1}{3} X_{12},\nonumber \\
&& [X_{1112}, X_{2222}] = 4 \, X_{112222} - \frac{1}{3} X_{22}, \nonumber \\
&& [X_{1122}, X_{1222}] = 8 \, X_{112222} + X_{22}, \nonumber \\
&& [X_{1122}, X_{2222}] = 10 \, X_{122222}, \qquad [X_{1222}, X_{2222}] = 20 \,  X_{222222} \ .
\nonumber
\end{eqnarray}

The spin-2 generators are orthogonal to all other spin-$s$ generators for $s > 2$.  Their scalar products among themselves are given by
\begin{eqnarray}
(X_{1111}, X_{2222}) = \frac{1}{12} , \qquad (X_{1112}, X_{2221}) = - \frac{1}{3}, \qquad
(X_{1122}, X_{2211}) = \frac{1}{8}
\nonumber
\end{eqnarray}
while all other scalar products are zero.

\section{Truncation to $sl(3,\re)$}
\setcounter{equation}{0}
\label{AppendixB}

The spin-1 and spin-2 generators $\{X^{(2)}, X^{(4)} \}$ span an 8-dimensional space. This is not a subalgebra since the brackets of two spin-2 generators contain terms involving spin 3-generators $x^{(6)}$.  If one forces these terms to zero by hand, the new brackets between spin-1 and spin-2 generators defined in this way close. In general, there is no guarantee that they fulfill the Jacobi identity since the brackets of the spin-3 generators with the spin-2 generators contain spin-2 generators.

It turns out this truncation is special, as the Jacobi identities are fulfilled. Notice that such a feature does not extend straightforwardly to truncation to higher levels. The Lie algebra defined by the truncation to spin-1 and spin-2 is actually isomorphic to $sl(3,\re)$. The easiest way to see this is to exhibit the isomorphism by expressing the Chevalley-Serre generators $\{H_1, H_2, E_1, E_2, F_1, F_2\}$ of $sl(3,\re)$ in terms of the $\{X^{(2)}, X^{(4)} \}$ generators.  We define
\begin{eqnarray}
&&H_1 =  -\frac{1}{4} \left(X_{12} + 2 \sqrt{3} X_{1122} \right), \; \; \; H_2 = -\frac{1}{4} \left(X_{12} - 2 \sqrt{3} X_{1122} \right), \nonumber \\
&&E_1 =  + \frac{1}{2} \left(X_{11} + \sqrt{3} X_{1112} \right), \; \; \; E_2 = + \frac{1}{2} \left(X_{11} - \sqrt{3} X_{1122} \right), \nonumber \\
&&F_1 =  -\frac{1}{4} \left(X_{22} +  \sqrt{3} X_{1222} \right), \; \; \; F_2 = -\frac{1}{4} \left(X_{22} -  \sqrt{3} X_{1222} \right), \nonumber \\
&& E_3 = - \sqrt{3} X_{1111}, \qquad F_3 = \frac{\sqrt{3}}{2} X_{2222}
\nonumber
\end{eqnarray}
The mapping
\begin{eqnarray}
&& H_1 \rightarrow h_1, \qquad H_2 \rightarrow h_2 \nonumber \\
&& E_1 \rightarrow e_1, \qquad E_2 \rightarrow e_2, \qquad E_3 \rightarrow [e_1,e_2],
\nonumber \\
&& F_1 \rightarrow f_1, \qquad F_2 \rightarrow f_2, \qquad F_3 \rightarrow [f_1,f_2]
\nonumber
\end{eqnarray}
preserves the bracket and is the isomorphism we are after.

\vspace{.5cm}
\noindent
{\bf Note added}

After completing this work, we were informed by Andrea Campoleoni, Stefan Fredenhagen, Stefan Pfenninger
and Stefan Theisen that they considered asymptotic symmetries for finite component, higher spin models in three dimensions based on $sl(N, \re)$ (or other real forms of $sl(N, {\mathbb{C}})$) and obtained the $W_N$ algebra.
Our result treats the infinite tower of higher spin fields on equal footing from the outset and directly obtains the nonlinearly realized, centrally extended $W_\infty$ symmetry algebra.

\end{document}